\documentclass[aps,11pt,english,tightenlinesletterpaper,superscriptaddress,secnumarabic,nofootinbib]{revtex4}

\usepackage{amsmath,amssymb}
\usepackage[T1]{fontenc}
\usepackage[latin9]{inputenc}
\usepackage[letterpaper]{geometry}
\usepackage{amssymb}
\usepackage{esint}
\usepackage{color}
\usepackage{graphicx}
\usepackage{verbatim}
\usepackage{slashed} 

\makeatother

\usepackage{babel}

\usepackage{hyperref}
\hypersetup{colorlinks=true,citecolor=blue}

\begin{document}

\title{Gravitational Fermion Production in Inflationary Cosmology}

\author{Daniel J.~H.~Chung, Lisa L.~Everett, Hojin Yoo, and Peng Zhou\\ \vspace{-0.05in}
Department of Physics, University of Wisconsin, Madison,
  WI 53706, USA}

\begin{abstract}
We revisit the gravitational production of massive Dirac fermions in
inflationary cosmology with a focus on clarifying the analytic
computation of the particle number density in both the large and the
small mass regimes.  For the case in which the masses of the
gravitationally produced fermions are small compared to the Hubble
expansion rate at the end of inflation, we obtain a universal result
for the number density that is nearly independent of the details of
the inflationary model.  The result is identical to the case of
conformally coupled scalars up to an overall multiplicative factor of
order unity for reasons other than just counting the fermionic degrees
of freedom.
\end{abstract}
\maketitle

\section{Introduction}

Gravitational particle production (as reviewed e.g~in
\cite{Birrell:1982ix,DeWitt1975}) and string production (see
e.g.~\cite{Lawrence:1995ct,Gubser:2003vk,Turok:2004gb,Friess:2004zk,Tolley:2005ak,Cremonini:2006sx,Das:2006dr,Feng:2007qk})
are generic phenomena for quantum fields in a curved spacetime
background and are analogs of particle creation in strong electric
fields (see e.g.~\cite{Schwinger1951,Brezin:1970xf}).  In the case of
Friedmann-Robertson-Walker (FRW) cosmology without inflation, it was
found \cite{Audretsch1978,Mamaev1978,Mamaev1975,Parker1971,Parker1969}
that the production of fermion and conformally coupled scalar fields
near the radiation dominated (RD) universe singularity occurs when the
particle masses $m$ are comparable to the Hubble expansion rate $H$,
with a number density $n\sim m^{3}$ that dilutes as $a^{-3}$ due to
expansion.  The fractional relic density of these particles at the
time of radiation-matter equality is
$\Omega_{X}\sim(m_{X}/10^{9}\mbox{GeV})^{5/2}$ \cite{Kuzmin}.  Hence,
the requirement of $\Omega_{X}<1$ puts an upper bound of $10^{9}\,
\mbox{GeV}$ on the stable particle mass.\footnote{Physics quite
  similar to this is reported in
  \cite{Anischenko:2009va,PhysRevD.64.023519}.}

In contrast, in inflationary cosmology the previously unbounded rapid
growth of $H$ as one moves backward in time towards the RD singularity
is replaced by a nearly constant $H_{e}$ during the quasi-de Sitter
(dS) era.  In such cases, the possibility of superheavy dark matter in
a wide range of masses including $m>H_e$ was emphasized in
\cite{Chung:1998zb,Kuzmin:1998kk}.  In fact, natural superheavy dark
matter candidates existed in the context of string phenomenology
before the gravitational production mechanism was appreciated
\cite{Ellis:1990iu,Benakli:1998ut}.  Furthermore, many extensions of
the Standard Model also possess superheavy dark matter candidates (see
e.g.~\cite{Kusenko:1997si,Han:1998pa,Dvali:1999tq,Hamaguchi:1999cv,Coriano:2001mg,Cheng:2002iz,Shiu:2003ta,Berezinsky:2008bg,Kephart:2001ix,Kephart:2006zd}),
which can have interesting astrophysical implications (see
e.g.~\cite{Coriano:2001mg,Barbot:2002gt,Albuquerque:2003ei,Taoso:2007qk,Bovy:2008gh,Albuquerque:2010bt}).
In such contexts, analytic relic density formulae have been computed
in the heavy and the light mass regimes for conformally coupled
scalars \cite{Chung2003,Chung:2001cb}).

In this work, we turn our attention to the gravitational particle
production of long-lived Dirac fermions in inflationary cosmology.
Gravitational particle production of Dirac fermions has been studied
numerically within the context of specific chaotic inflationary models
\cite{Kuzmin:1998kk}.  Our purpose is to clarify the analytic
computation and to derive a universal result for the light mass
scenario that is nearly independent of the details of the inflationary
model.  Our result is identical up to an overall $O(1)$ multiplicative
factor to that obtained for conformally coupled light scalar fields in
\cite{Chung:2001cb}, despite the fact that the Dirac structure naively
imposes a different spectral (momentum scaling) property on the
equations governing the particle production.\footnote{Although the aim
  of \cite{Chung:2001cb} is to consider a hybrid inflationary
  scenario, it also contains a universal result, equation (44),
  applicable to generic inflationary scenarios. There is also a
  misprint in \cite{Chung:2001cb} in stating that the situation is for
  minimal coupling rather than for conformal coupling.}  In comparison
to the conformally coupled scalar case, no special non-renormalizable
coupling to gravity nor possibility of tadpole instabilities concern
the fermionic scenario in the light mass limit because the fermion
kinetic operator is conformally invariant and fermions cannot obtain a
nonvanishing vacuum expectation value.

We also derive the particle production spectrum for the heavy mass
scenario and find it to be identical to the result of \cite{Chung2003}
(again up to an $O(1)$ multiplicative constant) despite a different
momentum dependence of the starting point of the equations.  As
expected, the heavy mass number density falls off exponentially.  In
contrast with the light mass limit, this case is sensitive to the
details of the transition out of the inflationary era.  To emphasize
the simplicity and the novel analytic arguments of the light mass
scenario, we relegate
the heavy mass results to an appendix.

It should be noted that the production of fermions in inflationary
cosmology has been extensively considered during the recent past, but
most analyses have focused on the non-gravitational interactions.  For
example,
\cite{Garbrecht:2002pd,PhysRevD.57.6003,Bassett:2001jg,Peloso:2000hy,GarciaBellido:2000dc,PhysRevD.58.125013,Greene:1998nh,Dolgov:1989us}
focused on both numerical and analytic analyses of fermion production
during preheating.  \cite{Chung:1999ve} considered the production
effects when the fermion mass passes through a zero during the
quasi-dS phase.  The effects of radiative corrections that modify the fermion
dispersion relationship and its connection to particle production were
considered in \cite{PhysRevD.73.064036}.  Gravitino production
has also been considered by many authors (see
e.g.~\cite{Maroto:1999ch,Kallosh:1999jj,Giudice:1999am,Nilles:2001fg,Nilles:2001ry,Kawasaki:2006hm}).
The main thrust of this paper differs in that it focuses on the
minimal gravitational coupling and derives a simple bound analogous to
Eq.~(44) of \cite{Chung:2001cb}.  Indeed, our results 
will aid in future investigations similar to \cite{Berezinsky:2008bg}
which would benefit from a more accurate simple analytic estimate of
the dark matter abundance.

The outline of this work is as follows. In Sec.~2, we discuss the
intuition behind the general formalism for the gravitational
production of massive Dirac fermions in curved spacetime.  In Sec.~3,
we discuss the generic features of the spectrum and derive the main
result of this paper, which is that for a given mode with comoving
wave number $k$, the Bogoliubov coefficient magnitude $|\beta_{k}|^2
\sim O(1/2)$ if $H(\eta)>m$ when $k/a(\eta)\sim m$.  We test this
analytic result within a toy inflationary model in Sec.~4, and then
discuss the dependence on reheating and the implications for the relic
density in Sec.~5.  Finally, in Sec.~6 we summarize our results and
present our conclusions.  Appendix A contains a collection of useful
results for fermionic Bogoliubov transformation computations.
Appendix B contains a complementary argument (which relies more on the
spinorial picture of the fermions) for the universality of the
Bogoliubov coefficient in the light mass region.  Appendix C contains
the particle density spectrum for the heavy mass limit.

\section{Fermion Particle Production: Background and Intuition}
To compute the particle production of Dirac fermions in curved
spacetime, we follow the standard procedure as outlined for example in
\cite{Birrell:1982ix,DeWitt1975} to calculate the Bogoliubov
coefficient $\beta_k$ between the in-vacuum corresponding to the
inflationary adiabatic vacuum and the out-vacuum corresponding to the
adiabatic vacuum defined at post-inflationary times.  The details of
this formalism and our conventions are presented in Appendix A, with
the expression for $\beta_k$ given in Eq.~(\ref{eq:betak}).

However, to obtain a better intuitive picture of the particle
production mechanism, here we present general physical arguments
regarding the expected features of the spectrum.  We begin by
considering a Dirac fermion field $\Psi$ described by
\begin{equation}
\mathcal{L}=i\bar{\Psi}\gamma^{\mu}\nabla_{\mu}\Psi-m\bar{\Psi}\Psi \label{eq:DiracLagrangian}
\end{equation}
minimally coupled to gravity. As the action $S=\int
d^{4}x\sqrt{g}\mathcal{L}$ is conformally invariant in the
$\{m\rightarrow0,\,\hbar\rightarrow0\}$ limit (with $\delta
g_{\mu\nu}(x)=-2\sigma(x)g_{\mu\nu}(x)$), physical quantities are
necessarily independent of the FRW scale factor $a$ to leading order
in $\hbar$.  Hence, the leading $\hbar$ order Bogoliubov coefficient
$\beta_k$ is zero in the $ma/k\rightarrow0$ limit, since it is the
metric that drives the particle production ({\it i.e.}, it plays the
role of the electric field in the analogy of particle creation by
strong electric fields).  This implies that particle production can
only occur in significant quantities for non-relativistic
modes.\footnote{We neglect possible conformal symmetry breaking
  effects associated with preheating \cite{Bassett:2001jg}.  In that
  sense, there is a mild implicit model dependence here.}

We next point out that the Dirac equation with a time-dependent mass
term results in mixing between positive and negative frequency modes,
similar to the case of the conformally coupled Klein-Gordon system
with a time-dependent mass.  To see this explicitly, consider the
Dirac equation for the spinor mode functions $u_{A,B}$ that follows
from Eq.~(\ref{eq:DiracLagrangian}):
\begin{equation}
i\partial_{\eta}\left(\begin{array}{c}
u_{A}\\
u_{B}
\end{array}\right)=\left(\begin{array}{cc}
am & k\\
k & -am
\end{array}\right)\left(\begin{array}{c}
u_{A}\\
u_{B}
\end{array}\right),\label{eq:timedepmodefunc}
\end{equation}
which is our Eq.~(\ref{eq:Dirac Equation}) from Appendix A.
Here $u_{A,B}$ span the complete solution space (they contain both
approximate positive and negative frequency solutions in the adiabatic
regime).  Here we are working in conformal time, which is related
to the comoving observer's proper time via $dt\equiv a(\eta)d\eta$.
From Eq.~(\ref{eq:timedepmodefunc}), we see that the rotation matrix
that diagonalizes the right hand side is a function of the
time-dependent quantity $am$. Hence, the Dirac equation \emph{as a
  function of time} mixes approximate positive and negative frequency
solutions leading to non-vanishing particle production.

To estimate the Bogoliubov coefficient, we can compute the effects of
the time-dependent mixing matrix $\mathcal{U}\in O(2)$ as follows.
We begin by inserting $1=\mathcal{U}^{T}\mathcal{U}$ into
Eq.~(\ref{eq:timedepmodefunc}) to obtain
\begin{eqnarray}
i\mathcal{U}\partial_{\eta}\left[\mathcal{U}^{T}\mathcal{U}\left(\begin{array}{c}
u_{A}\\
u_{B}
\end{array}\right)\right] & = & \mathcal{U}\left(\begin{array}{cc}
am & k\\
k & -am
\end{array}\right)\mathcal{U}^{T}\mathcal{U}\left(\begin{array}{c}
u_{A}\\
u_{B}
\end{array}\right)\\
\implies i\mathcal{U}\partial_{\eta}\mathcal{U}^{T}\left(\begin{array}{c}
u_{A}'\\
u_{B}'
\end{array}\right)+i\partial_{\eta}\left(\begin{array}{c}
u_{A}'\\
u_{B}'
\end{array}\right) & = & \left(\begin{array}{cc}
\sqrt{k^{2}+m^{2}a^{2}} & 0\\
0 & -\sqrt{k^{2}+m^{2}a^{2}}
\end{array}\right)\left(\begin{array}{c}
u_{A}'\\
u_{B}'
\end{array}\right),\label{eq:matrixform}
\end{eqnarray}
in which the primed basis is defined to be  
\begin{equation}
\left(\begin{array}{c}
u_{A}'\\
u_{B}'
\end{array}\right)\equiv\mathcal{U}\left(\begin{array}{c}
u_{A}\\
u_{B}
\end{array}\right).
\end{equation}
The Dirac equation is diagonal in the primed basis except for the
appearance of the mixing term
\begin{equation}
\mathcal{U}\partial_{\eta}\mathcal{U}^{T}=\frac{a}{2}\left(\begin{array}{cc}
0 & \frac{mHk_{p}}{k_{p}^{2}+m^{2}}\\
-\frac{mHk_{p}}{k_{p}^{2}+m^{2}} & 0
\end{array}\right),
\end{equation}
 with $k_{p}\equiv k/a$. From this result, we see that during inflation the mixing term approximately vanishes for a fixed comoving wave number $k$ as $a\rightarrow0$, while after inflation it is the largest when $H$ is the largest.  Using this result, it is straightforward to show that the Bogoliubov coefficients due to mixing take the following form:
\begin{equation}
\beta_{k}^{\mbox{mix}}\sim\int dt\frac{mk_{p}}{k_{p}^{2}+m^{2}}He^{-2i\int dt\, \omega_{k}}, \label{eq:bogoliubov}
\end{equation}
in which $\omega_{k}=\sqrt{k_{p}^{2}+m^{2}}$. One may still ask whether there are any other sources of positive
and negative frequency mixing since the diagonal terms of Eq.~(\ref{eq:matrixform}) are time dependent, just as conformally coupled scalar fields contain $\omega^{2}=k^{2}+m^{2}a^{2}$ in their mode equations. The answer is no \emph{if} the fermionic particles are \emph{defined} as modes that exactly satisfy the condition  
\begin{equation}
i\partial_{\eta}\left(\begin{array}{c}
u_{A}'\\
u_{B}'
\end{array}\right)=\left(\begin{array}{cc}
\sqrt{k^{2}+m^{2}a^{2}} & 0\\
0 & -\sqrt{k^{2}+m^{2}a^{2}}
\end{array}\right)\left(\begin{array}{c}
u_{A}'\\
u_{B}'
\end{array}\right).
\end{equation}
For example, the adiabatic vacuum positive frequency modes are defined to be 
\begin{equation}
\left(\begin{array}{c}
u_{A}'\\
u_{B}'
\end{array}\right)
\propto\left(\begin{array}{c}
1\\
0
\end{array}\right)e^{-i\int dt\sqrt{\frac{k^{2}}{a^{2}}+m^{2}}}.\label{eq:zerothadiabatic}
\end{equation}
Eq.~(\ref{eq:zerothadiabatic}) corresponds to a zeroth order adiabatic
vacuum in which the adiabaticity parameter $\epsilon_{A}$ is defined as
\begin{equation}
\epsilon_{A}\equiv\frac{mHk_{p}}{\left(k_{p}^{2}+m^{2}\right)^{3/2}},
\end{equation}
in accordance with the usual conventions
\cite{Birrell:1982ix,Parker1969,Chung:1998zb,Chung:2003wn}.  This
parameter vanishes in the asymptotically far past (near when the in-vacuum
is defined) and in the far future (near when the out-vacuum is defined).
Eq.~(\ref{eq:zerothadiabatic}) coincides with
\begin{equation}
\left(\begin{array}{c}
u_{A}\\
u_{B}
\end{array}\right)\rightarrow\left(\begin{array}{c}
\sqrt{\frac{\omega+am}{2\omega}}\\
\sqrt{\frac{\omega-am}{2\omega}}
\end{array}\right)e^{-i\int^{\eta}d\eta'\omega}
\end{equation}
 in the basis of Eq.~(\ref{eq:timedepmodefunc}).

 To summarize, the zeroth adiabatic order vacuum Bogoliubov coefficient is approximately given by Eq.~(\ref{eq:bogoliubov}).  We see that compared to the conformally coupled bosonic case (see e.g.~\cite{Chung2003}), the long wavelength fermionic particle production is suppressed due to the appearance of $k_p$ in the numerator.

\section{\label{sec:genericlightmass} Light Mass Case and Generic Features of the Spectrum}

In this section, we present a universal result for the spectrum in the light mass scenario that is nearly independent of the details of the inflationary model.   We will show that under a specific set of conditions, the Bogoliubov spectral amplitude (evaluated with observable particle state basis defined at time $t$) takes the approximate form
\begin{equation}
|\beta_k(t)|^2 \sim O(1/2).
\label{eq:betaksqsimo1}
\end{equation} 
An alternate argument emphasizing more of the spinorial nature of the
fermions is presented in Appendix B.

For Eq.~(\ref{eq:betaksqsimo1}) to hold generically, the following conditions must be simultaneously satisfied.  The fermions that are produced must be light (to be made precise below).   After the end of inflation, the modes that are produced must become non-relativistic during the time when the expansion rate is the dominant mass scale.  Finally, $t$ must be at a time when the particles with momentum $k_p=k/a$ are non-relativistic.

To see this more explicitly, we note that because relativistic modes are approximately conformally invariant,
modes that can be significantly produced by the FRW expansion satisfy 
$k_{p}\lesssim m$,
 where $\vec{k}_{p}$ is the physical momentum during the time period of interest. Furthermore, during the time that
 $k_{p}\lesssim m$, Eq.~(\ref{eq:bogoliubov}) takes the form
\begin{eqnarray}
\beta_{k}(t) & \sim & \int^t dt'\frac{k_{p}(t')}{m}H(t')e^{-2i\int^{t'} dt''\omega_{k}(t'')}.\label{eq:bogoliubovintegrand}
\end{eqnarray}
Let us consider Eq.~(\ref{eq:bogoliubovintegrand}) for the time period with $H(t')>m$, such that $|\dot{H}|\sim H^2 > \omega_k^2 $.  Here we take $k$ to be consistent with $k_p\lesssim m$;  more precisely, $m a(t)>k>ma(t_i)$, where $t_i$ is the time when the initial vacuum is defined, which is typically at the beginning of inflation.  In this regime, the largest contribution to $\beta_k$ arises from the time $t_{\mbox{max}}(k)$ when $k_p=k/a$ is
at its largest while remaining non-relativistic ($k/a(t_{\mbox{max}})=m$).
\footnote{The condition $k>ma(t_i)$ comes from the requirement of setting the adiabatic vacuum condition, which only applies for modes with subhorizon wavelengths.}
Hence, in this case Eq.~(\ref{eq:bogoliubovintegrand}) results in 
\begin{equation}
\beta_{k}(t) \sim O\left(\frac{k/a(t_{\mbox{max}}(k))}{m}\right),\label{eq:betafirstapprox}
\end{equation}
which is indeed of $O(1)$.




We note that Eq.~(\ref{eq:betafirstapprox}) is independent of $H$, indicating an insensitivity to the details of the inflationary model.  This holds as long as the dominant contribution to Eq.~(\ref{eq:bogoliubovintegrand}) arises from the time period with $H(t')/m >1$.  The condition $H(t')/m>1$ fails if $t'>t_m$ where $H(t_m)=m$. Thus, there is a mild inflationary model dependence of $m/H_e <1$, where $H_e\equiv H(t_e)$ is the expansion rate at the end of
inflation.  As there is a general restriction that $|\beta_k|^2 < 1$ from quantization conditions, $O(1)$ here must mean a number
less than unity.\footnote{The Bogoliubov coefficients satisfy
  $|\alpha_{\vec{k},s}|^2+|\beta_{-\vec{k},s}|^2=1$, while
  Eq.~(\ref{eq:betafirstapprox}) effectively neglects this
  constraint.}  To remind ourselves of this fact, we
will refer to this $O(1)<1$ number as $O(1/\sqrt{2})$.  Putting all the
conditions together with Eq.~(\ref{eq:betafirstapprox}), we find
\begin{equation}
|\beta_{k}(t)|^2 \sim O(1/2)\,\,\,\,\,\,\mbox{ for }
t_m>t_{\mbox{max}}(k)>t_{e}\mbox{ and } t>t_{\mbox{max}}(k), \label{eq:mainresult}
\end{equation}
A more explicit restriction
 on the $k$ values corresponding to the requirements of  Eq.~(\ref{eq:mainresult}) can be written as follows:
\begin{equation}
m a(t_m)\gtrsim k > m a(t_i) \mbox{ and } ma(t)\gtrsim k.
\label{eq:kvalues}
\end{equation}
Eqs.~(\ref{eq:mainresult}) and (\ref{eq:kvalues}) are the main results
of this section.
 
For modes with $k> m a(t_m)$, $|\beta_{k}|^{2}$ is smaller since Eq.~(\ref{eq:bogoliubovintegrand}) is suppressed
by an additional factor of $H/m$.  The exact high $k$ behavior of
$\beta_k$ is sensitive to the adiabatic order of the vacuum boundary
condition as well as the details of the scale factor during the
transition out of the quasi-dS era.  However, what is generic is that the
spectral contribution to the particle density no longer grows
appreciably when $k> m a(t_m)$.  Hence, we define the critical momentum
$k_*\equiv m a(t_m)$, which in terms of the momentum at the end of inflation is given by 
\begin{equation}
k_*/a_e = (H_e/m)^{2/n_a} m,
\label{eq:criticalmom}
\end{equation}
where we parameterized the energy density after the end of inflation
as $\rho\propto a^{-n_a}$.
Integrating over $d^{3}k/(2\pi a)^{3}$ to obtain the energy density of the fermions, we can for an order of magnitude
estimate introduce a step function $\Theta(k_{*}-k)$ as follows:
\begin{equation}
\rho_{\Psi}(t)\sim4\times\frac{m}{4\pi^{2}}\frac{1}{a^{3}}\int dkk^{2}\Theta(k_{*}-k),\,\,\,\,\,\,\,\,\,\,\,\,\, t_{\mbox{max}}(ma(t_{i}))\ll t_{m}<t.\label{eq:densityintegral}
\end{equation}
 Assuming that the  lower limits of Eq.~(\ref{eq:densityintegral}) make a negligible contribution, we obtain 
\begin{equation}
\rho_{\Psi}(t)\sim4\times\frac{m^{4}}{12\pi^{2}}\left(\frac{a(t_{m})}{a(t)}\right)^{3},\label{eq:mainresult2}
\end{equation}
which contains the mild inflationary scenario dependence of $m<H_{e}$.  As we will see in Sec.~\ref{sec:reheatingdep}, a stronger inflationary model dependence arises from the dilution factor $a(t_{m})/a(t)$, which typically is a function of the reheating temperature.

\section{\label{sec:instantaneous}Example of Fermion Production in a Toy Inflationary Model}

To test the analytic estimation of Sec.~\ref{sec:genericlightmass}, we now numerically compute the particle production in a toy inflationary model with instantaneous reheating occurs ({\it i.e.}, in which the~quasi-dS phase connects instantaneously to the RD phase).  
As is well known, such non-analytic models have unphysical large momentum behavior \cite{Birrell:1982ix}, which for our purposes can be dealt with simply by cutting off the integration of the spectrum.  We find there is an upper bound on the fermion mass if $m<H_{e}$ during inflation, similar to the case of fermion production in pure RD cosmology \cite{Kuzmin}.  We will turn to the more realistic
case in which the inflationary era exits to a transient pressureless era during reheating in Sec.~\ref{sec:reheatingdep}.

Let us consider a background spacetime which is initially dS with a
Hubble constant $H_{e}$ that is followed by RD spacetime.  Although
the junction between the dS and RD eras is instantaneous, the scale
factor $a(t)$ and the Hubble rate $H(t)$ are continuous across the
junction.  In particular, if we set the junction time at the conformal
time $\eta=0$ and we set the scale factor at the junction time to be
$a_{e}$, the scale factor and Hubble rates can be written as
\begin{equation}
a(\eta)=\begin{cases}
\left (\left ( \frac{1}{a_{e}H_{e}}-\eta \right)H_{e} \right)^{-1} & \eta\le0\;(\mbox{dS})\\
a_{e}^{2}H_{e}\left (\eta+\frac{1}{a_{e}H_{e}}\right ) & \eta>0\;(\mbox{RD}),
\end{cases} \qquad 
H(\eta)=\begin{cases}
H_{e} & \eta\le0 \;(\mbox{dS})\\
H_{e}\left(\frac{a_{e}}{a(\eta)}\right)^{2} & \eta>0\;(\mbox{RD}),
\end{cases}
\end{equation}
indicating that the leading discontinuity in $a$ occurs at second order in the conformal time derivative.

To compute $\beta_k$ using Eq.~(\ref{eq:betak}), it is necessary to fix the boundary conditions for the in-modes and the out-modes.   For the in-modes, we require that in the infinite past, when a certain given mode's
wavelength is within the horizon radius, its mode function must agree with the flat space positive frequency mode function.  In other words, as $a(\eta)\rightarrow0$, 
\begin{equation}
\left(\begin{array}{c}
u_{A}\\
u_{B}
\end{array}\right)_{k,\eta}^{in}\rightarrow\left(\begin{array}{c}
\sqrt{\frac{\omega+a(\eta)m}{2\omega}}\\
\sqrt{\frac{\omega-a(\eta)m}{2\omega}}
\end{array}\right)e^{-i\int^{\eta}\omega(\eta')d\eta'}.\label{eq:in-BC}
\end{equation}
The in-modes' analytic expressions during the dS era thus take the form
\begin{eqnarray}
\left(\begin{array}{c}
u_{A}\\
u_{B}
\end{array}\right)_{k,\eta}^{in} & = & \left(\begin{array}{c}
\sqrt{\frac{\pi}{4}(\frac{k}{aH_{e}})}e^{i\frac{\pi}{2}(1-i\frac{m}{H_{e}})}H_{\frac{1}{2}-i\frac{m}{H_{e}}}^{(1)}(\frac{k}{aH_{e}})\\
\sqrt{\frac{\pi}{4}(\frac{k}{aH_{e}})}e^{i\frac{\pi}{2}(1+i\frac{m}{H})}H_{\frac{1}{2}+i\frac{m}{H_{e}}}^{(1)}(\frac{k}{aH_{e}})
\end{array}\right)
\end{eqnarray}
where $H_\nu^{(1)}$ are Hankel functions of the first kind.
Similarly, for the out-modes, as $k/a>H(\eta)$ in the RD era,
we require the mode functions to agree with the flat
space positive frequency mode functions, {\it i.e.}, as  $a(\eta)\rightarrow+\infty$,
\begin{equation}
\left(\begin{array}{c}
u_{A}\\
u_{B}
\end{array}\right)_{k,\eta}^{out}\rightarrow\left(\begin{array}{c}
\sqrt{\frac{\omega+a(\eta)m}{2\omega}}\\
\sqrt{\frac{\omega-a(\eta)m}{2\omega}}
\end{array}\right)e^{-i\int^{\eta}\omega(\eta')d\eta'}.
\end{equation}
The out-mode analytic expressions during the RD era are given by 
\begin{equation}
\left(\begin{array}{c}
u_{A}\\
u_{B}
\end{array}\right)_{k,\eta}^{out}=\left(\begin{array}{c}
e^{-\frac{\pi}{4}C}D_{-iC}(e^{i\pi/4}\sqrt{\frac{2m}{H(\eta)}})\\
\sqrt{C}e^{-\frac{\pi}{4}C+i\frac{\pi}{4}}D_{-iC-1}(e^{i\pi/4}\sqrt{\frac{2m}{H(\eta)}})
\end{array}\right),
\end{equation}
in which $C\equiv (k^{2}/a_{e}^2)/(2mH_{e})$ characterizes the ratio of the 
momentum to the dynamical mass scale and the $D_{v}(x)$ are parabolic
cylinder functions.  

\begin{figure}
\centering{}\includegraphics[width=0.7\paperwidth]{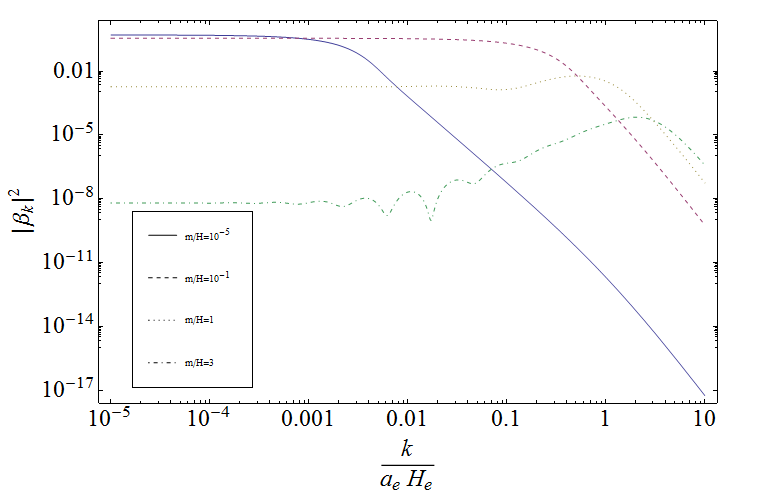}\caption{\label{fig:Bogoliubov-coefficients} The Bogoliubov coefficient amplitude $|\beta_{k}|^{2}$ as a function of $k/(a_eH_e)$ for various ratios of the fermion mass to the Hubble expansion rate during the dS era. }
\end{figure}

The numerical results for $|\beta_{k}|^{2}$ are shown as a function of $k/(a_e H_e)$ for various choices of the fermion masses in Fig.~\ref{fig:Bogoliubov-coefficients}.
From these results, we first note that it can be determined that for heavy masses $m>H_{e}$, e.g.~$m/H_{e}=1 $ or 3, the
infrared end of the spectrum behaves as $|\beta_{k}|^{2}\sim (1+\exp(2\pi m/H_e))^{-1}$.  Further details of the heavy mass case are given in Appendix~\ref{sec:heavymasscase}.  As the heavy mass situation is likely to be more sensitive to the abrupt transition approximation made in this section, we restrict our attention here to the light mass case in which $m<H_{e}$.

For the light mass case (e.g.~$m/H_{e}=10^{-5}$ in Fig.~(\ref{fig:Bogoliubov-coefficients})),
we can see there are three ranges of $k$ that each have qualitatively different behavior.  For  $k/a_{e}>H_{e}$, the modes are still inside the horizon at the end of inflation, and the spectrum falls off as $|\beta_{k}|^{2}\propto k^{-6}$.
In contrast, for $\sqrt{mH_{e}}<k/a_{e}<H_{e}$, the modes are outside of the horizon at the end of inflation and remain relativistic at the time when $m=H(\eta)$ during RD.  In this case, the spectrum falls off as $|\beta_{k}|^{2}\propto k^{-4}$.   Finally, for $k/a_{e}<\sqrt{mH_{e}}$, the modes are outside the horizon at the end of inflation and have become non-relativistic before $m=H(\eta)$ during RD.  This results in a constant spectrum of 
  $|\beta_{k}|^{2}\approx\frac{1}{2}$, in agreement with the  results of Sec.~\ref{sec:genericlightmass}.  
Generically, if the scale factor $a(\eta)$ is sufficiently continuous \cite{Parker1971,Chung:2003wn}, the spectrum will fall off in the ultraviolet region faster than $k^{-3}$, such that the total number density $n\sim\int d^{3}k|\beta_{k}|^{2}$ is finite. The majority of the
contribution arises from the region in which $k/a_{e}<\sqrt{mH_{e}}$ where $|\beta_{k}|^{2}\approx\frac{1}{2}$, as anticipated in
Sec.~\ref{sec:genericlightmass}.
The number density for particle masses in the range of $m<0.1H_{e}$ is numerically determined to be (recall that $\eta_{m}$ is defined by $H(\eta_m)=m$)
\begin{equation}
n(\eta)=4\times0.005m^{3}\left(\frac{a(\eta_{m})}{a(\eta)}\right)^{3},
\end{equation}
which again agrees with the analytic estimate of  
Eq.~(\ref{eq:mainresult2}).

\section{\label{sec:reheatingdep} Inflationary Reheating Dependence}
We now consider the more realistic situation in which there is a
smooth transition region between the dS and RD phases.  When inflation
ends, there is typically a period of coherent oscillations
($a_{e}<a<a_{\mbox{rh}}$) during which the equation of state is close
to zero (see e.g.~\cite{Kolb:1990vq,Lyth:1998xn,Mazumdar:2011zd}).
During that period, the expansion rate behaves as $H\propto a^{-3/2}$
and not $a^{-2}$ as during RD.  This difference will lead to an
effective dilution of the dark matter particles by the time RD is
reached.  More precisely, the fermion number density will be diluted
as $1/a^{3}$ as long as the fermion plus anti-fermion number is
approximately conserved.  As we will see below, the integrated
dilution is typically a function of the reheating temperature during
inflation.

Accounting for the dilution, in this section we estimate the relic
abundance of fermionic particles (fermions plus
anti-fermions).\footnote{This requires the fermion self-annihilation
  cross section rate to be smaller than the expansion rate throughout
  its history.  Such weak interactions generically can be achieved for
  sufficiently large particle masses \cite{Chung:1998zb}, which are
  allowed as long as the inflationary scale is sufficiently large.}
The dilution consideration breaks up
naturally into two cases: $a_{\mbox{rh}}>a(t_m)$ and
$a_{\mbox{rh}}<a(t_m)$.  The former case corresponds to the situation
in which the dominant particle production occurs during the reheating
period, while the latter case corresponds to
the complementary situation, which we will see below is unlikely to be
physically important.

Let us begin with the case of $a_{\mbox{rh}}>a_{m}$, which corresponds
to
\begin{equation}
H_{e}\gg m>H_{\mbox{rh}}\sim\frac{\sqrt{g_{*}}}{3}\frac{T_{\mbox{rh}}^{2}}{M_{p}}=\left(\frac{T_{\mbox{rh}}}{10^{9}\mbox{ GeV}}\right)^{2}\left(\frac{g_{*}}{100}\right)^{1/2}\mbox{GeV},
\end{equation}
where $H_{\mbox{rh}}$ is the expansion rate at the time radiation
domination is achieved. In this case, we have 
\begin{equation}
\rho_{\Psi}(t_{\mbox{eq}}) \sim  0.03 m^{4}\left(\frac{H_{\mbox{rh}}}{m}\right)^{2}\left(\frac{a_{\mbox{rh}}}{a_{\mbox{eq}}}\right)^{3},
\end{equation}
in which we have used the fact that $H\propto a^{-3/2}$ during
reheating. We thus find the relic abundance today of fermionic
particles to be
\begin{equation}
\Omega_\psi h^2 \sim 3 \left(\frac{m}{10^{11}\mbox{
    GeV}}\right)^{2}\left(\frac{T_{\mbox{rh}}}{10^{9}\mbox{
    GeV}}\right).
\label{eq:morerealistic}
\end{equation}
This matches Eq.~(44) of \cite{Chung:2001cb} (up to a factor of order
of a few, part of which is expected from counting fermionic degrees of
freedom), which was computed in the context of conformally coupled
scalar fields.  The match is interesting because the analog of
Eq.~(\ref{eq:bogoliubov}) for the conformally coupled scalar field
case has a different $k/a$ dependence that converts into an effective
$m$ dependence due to the conformal invariance of the fermionic
kinetic term.  Eq.~(\ref{eq:morerealistic}) also agrees with the model
dependent numerical results of \cite{Kuzmin:1998kk} up to a factor of
10.  The related ratio of the fermion energy density to the radiation
energy density at matter radiation equality,
$\rho_{\Psi}(t_{\mbox{eq}})/\rho_{R}(t_{\mbox{eq}})$, is the same as
Eq.~(\ref{eq:morerealistic}) up to a factor of 10.

For the case with $a_{\mbox{rh}}<a_{m}$, we have
\begin{eqnarray}
\rho_{\Psi}(t_{\mbox{eq}}) & \sim & 0.03 m^{4}\left(\frac{a_{\mbox{m}}}{a_{\mbox{eq}}}\right)^{3},
\end{eqnarray}
 which leads to
\begin{equation}
\frac{\rho_{\Psi}(t_{\mbox{eq}})}{\rho_{R}(t_{\mbox{eq}})}\sim\left(\frac{m}{10^{8}\mbox{
    GeV}}\right)^{5/2} \left(\frac{g_*(t_m)}{100} \right)^{-1/4}\label{eq:originalbound}
\end{equation}
which up to an order of magnitude is $\Omega_\psi$.  However, since
this applies only for
\begin{equation}
m<\left(\frac{T_{\mbox{rh}}}{10^{9}\mbox{ GeV}}\right)^{2}\left(\frac{g_{*}}{100}\right)^{1/2}\mbox{GeV},
\end{equation}
the relic abundance is negligible in this case. For example, a
$m\sim1\, \mbox{GeV}$ benchmark point will render
$\Omega_{\Psi}\sim10^{-20}$.

\section{Conclusions}
In this paper, we revisited the gravitational production of massive
Dirac fermions in inflationary cosmology.  For the situation in which
the fermions are light compared to the Hubble expansion rate at the
end of inflation, we obtained the analytic result that the Bogoliubov
coefficient amplitude $\vert \beta_k(t)\vert^2 \sim 1/2$ if $H>m$ when
$k/a\sim m$, as summarized in Eqs.~(\ref{eq:mainresult}) and
(\ref{eq:kvalues}).  We used this result to compute the relic density
assuming that the gravitationally produced fermions are superheavy
dark matter particles.  In cases of phenomenological interest, the
dark matter relic abundance depends on the reheating temperature, as
given in Eq.~(\ref{eq:morerealistic}).  Up to a multiplicative overall
factor of $O(1)$, this result is identical to that obtained for
conformally coupled scalars in \cite{Chung:2001cb}.  In the case that
the fermions are heavy compared to the Hubble expansion rate at the
end of inflation, the relic abundance is given by
Eq.~(\ref{eq:heavymasscase}).

It is also of interest to consider the isocurvature behavior of the
gravitationally produced fermions in the case that they have suitable
long-range nongravitational interactions.  Work along these lines is
currently in progress \cite{ChungPengHojin}.

\begin{appendix}

\section{Formalism and Conventions}
Here we follow the strategy outlined in the classical review paper of DeWitt \cite{DeWitt1975}.   Consider the action of a four component
Dirac spinor in curved spacetime:
\begin{equation}
S=\int d^{4}x\sqrt{|g(x)|}\bar{\Psi}(i\gamma^{a}\nabla_{e_{a}}-m)\Psi\label{eq:action}
\end{equation}
in which the gamma matrices
$\gamma^{a}$ are chosen to be in the Dirac basis
\begin{equation}
\gamma^{0}=\left(\begin{array}{cc}
I & 0\\
0 & -I
\end{array}\right),\quad\gamma^{i}=\left(\begin{array}{cc}
0 & \sigma^{i}\\
-\sigma^{i} & 0
\end{array}\right)
\end{equation} 
to simplify the derivation of the second
order differential equation of the spinor mode functions.
Extremizing the action with respect to $\delta\bar{\Psi}$
and $\delta\Psi$ yields the equations of motion:
\begin{eqnarray}
(i\gamma^{\alpha}\nabla_{e_{\alpha}}-m)\Psi  =  0, \qquad \nabla_{e_{a}}\bar{\Psi}(-i\gamma^{a})-\bar{\Psi}m  =  0. \label{eq:EOM}
\end{eqnarray}
The solution space can be endowed with a scalar product as
\begin{equation}
(\Psi_{1},\Psi_{2})_{\Sigma}=\int d\Sigma n_{\mu}e_{a}^{\,\mu}\bar{\Psi}_{1}\gamma^{a}\Psi_{2}
\end{equation}
in which $\Sigma$ is an arbitrary space-like hypersurface, $d\Sigma$
is the volume 3-form on this hypersurface computed with the induced
metric, and $n_{\mu}$ is the future-pointing time-like unit vector
normal to $\Sigma$. The current conservation condition 
\begin{equation}
\nabla_{e_{a}}(\bar{\Psi}_{1}\gamma^{a}\Psi_{2})=0
\end{equation}
implies the integral in the scalar product is independent of the choice
of $\Sigma$.   The conjugation map can also be defined in the solution space as $Psi\mapsto-i\gamma^{2}\Psi^{*}$,
which induces a pairing in the solution space. 

Based on the scalar product and the conjugation map, 
one can construct an orthonormal basis for the solution space.  It can be written as $\{U_{i},V_{i}\equiv-i\gamma^{2}U_{i}^{*}\}$ ($i$ labels different solutions), with 
\begin{eqnarray}
(U_{i},U_{j})  =  \delta_{ij}, \qquad (U_{i},V_{j})  =  0.
\end{eqnarray}
The Heisenberg picture field operator $\Psi(x)$ can then be expanded in this basis as follows: 
\begin{eqnarray}
\Psi(x) & = & \sum_{i}a_{i}U_{i}+b_{i}^{\dagger}V_{i},
\end{eqnarray}
in which the canonical anticommutation relations imposed on equal-time surfaces and the orthonormality of the mode functions ensures that 
\begin{eqnarray}
\{a_{i},a_{j}^{\dagger}\}  =  \delta_{ij}, \qquad \{b_{i},b_{j}^{\dagger}\}  =  \delta_{ij}.
\end{eqnarray}
The vacuum state is defined by 
$a_{i}|{\rm vac}\rangle=b_{i}|{\rm vac}\rangle=0$.
The full Hilbert space can then be constructed as usual by applying the creation operators $a_i^\dagger$ and $b_i^\dagger$ to the vacuum state.


However, the choice of the orthonormal basis $\{U_{i},V_{i}\}$ is not unique. Consider a different orthonormal basis $\{\tilde{U}{}_{i},\tilde{V}{}_{i}\}$, which is related to the original basis as follows:
\begin{eqnarray}
\tilde{U}{}_{i}  =  \sum_{j}\alpha_{ij}U_{j}+\beta_{ij}V_{j}, \qquad \tilde{V}{}_{i}  =  \sum_{j}\alpha_{ij}^{*}V_{j}+\beta_{ij}^{*}U_{j}.
\end{eqnarray}
The Bogoliubov coefficients $\alpha_{ij}$ and $\beta_{ij}$ can be
extracted as 
\begin{eqnarray}
\beta_{ij}  =  (V_{j},\tilde{U}{}_{i}), \qquad 
\alpha_{ij}  =  (U_{j},\tilde{U}{}_{i})
\end{eqnarray}
Note that the orthonormality relation on $\{U_{i},V_{i}\}$ and $\{\tilde{U}{}_{i},\tilde{V}{}_{i}\}$
implies the following relation:
\begin{equation}
\left(\begin{array}{cc}
\alpha & \beta\\
\beta^{*} & \alpha^{*}
\end{array}\right)^{*}\left(\begin{array}{cc}
\alpha & \beta\\
\beta^{*} & \alpha^{*}
\end{array}\right)^{T}=\left(\begin{array}{cc}
I & 0\\
0 & I
\end{array}\right).
\end{equation}
Using $\Psi=\sum_{i}a_{i}U_{i}+b_{i}^{\dagger}V_{i}=\sum_{i}\tilde{a_{i}}\tilde{U_{i}}+\tilde{b}_{i}^{\dagger}\tilde{V_{i}}$, the following relation is obtained: 
\begin{equation}
\left(\begin{array}{c}
\tilde{a}\\
\tilde{b}^{\dagger}
\end{array}\right)=\left(\begin{array}{cc}
\alpha^{*} & \beta^{*}\\
\beta & \alpha
\end{array}\right)\left(\begin{array}{c}
a\\
b^{\dagger}
\end{array}\right).
\end{equation}
Hence, the two mode functions result in inequivalent vacua. To see this more explicitly, consider the
expectation value of the occupation number operator $\tilde{a_{i}}^{\dagger}\tilde{a_{i}}$ with respect to the vacuum defined using the  $a_{i},b_{i}$ operators:
\begin{eqnarray}
\langle {\rm vac}|\tilde{a_{i}}^{\dagger}\tilde{a_{i}}|{\rm vac}\rangle & = & \sum_{j}|\beta_{ij}|^{2}.
\end{eqnarray}
The vacuum state corresponding to one definition thus is an excited state in the other definition. 

We turn now to FRW spacetime, in which the metric is conformally flat:
\begin{equation}
ds^{2}=g_{\mu\nu}dx^{\mu}dx^{\nu}=a(x_{0})^{2}\eta_{\mu\nu}dx^{\mu}dx^{\nu}.
\end{equation}
Since the action of Eq.~(\ref{eq:action}) is covariant under Weyl transformations:
\begin{eqnarray}
g_{\mu\nu} =  \Omega^{2}(x)\tilde{g}_{\mu\nu},\,\,\Psi=\Omega(x)^{-3/2}\tilde{\psi}, \qquad    e_{a}^{\,\,\mu}=\Omega(x)^{-1}\tilde{e}_{a}^{\,\,\mu},
\end{eqnarray}
a  Weyl transformation with $\Omega(x)=a(x_{0})$ can be used to rewrite the action as follows:
\begin{equation}
S=\int d^{4}x\bar{\psi}(i\gamma^{\mu}\partial_{\mu}-a(\eta)m)\psi,
\end{equation}
where $\eta$ is the conformal time and $\psi$ is the rescaled
spinor field. The equation of motion now takes the form
\begin{equation}
(i\gamma^{\mu}\partial_{\mu}-a(\eta)m)\psi=0.\label{eq:rescaled_EOM}
\end{equation}
The solution space is spanned by the orthonormal basis $\{U_{\vec{k},r},V_{\vec{k},r}\}$,  which can be written as follows: %
\begin{eqnarray}
U_{\vec{k},r}(\eta,\vec{x}) & = & \frac{e^{i\vec{k}\cdot\vec{x}}}{(2\pi)^{3/2}}\left(\begin{array}{c}
u_{A,k,\eta}h_{\hat{k},r}\\
r\, u_{B,k,\eta}h_{\hat{k},r}
\end{array}\right)\\
 & \equiv & \frac{e^{i\vec{k}\cdot\vec{x}}}{(2\pi)^{3/2}}\left(\begin{array}{c}
u_{A,k,\eta}\\
r\, u_{B,k,\eta}
\end{array}\right)\otimes h_{\hat{k},r},\label{eq:mode-ansatz}
\end{eqnarray}
in which $\hat{k}$ is the unit vector in the $\vec{k}$ direction ($\hat{k}=\hat{e}_{z}$
if $\vec{k}=0$), and $h_{\hat{k},r}$ is a 2-component complex column
vector (called the helicity 2-spinor) that satisfies 
\begin{equation}
\hat{k}\cdot\vec{\sigma}h_{\hat{k},r}=rh_{\hat{k},r},\, r=\pm1.
\end{equation}
More concretely, if $\hat{k}=(\theta,\phi)$ in spherical coordinates,
then the normalization factor can be chosen such that 
\begin{equation}
h_{\hat{k},+1}\equiv\left(\begin{array}{c}
\cos\frac{\theta}{2}e^{-i\phi}\\
\sin\frac{\theta}{2}
\end{array}\right),h_{\hat{k},-1}\equiv\left(\begin{array}{c}
\sin\frac{\theta}{2}e^{-i\phi}\\
-\cos\frac{\theta}{2}
\end{array}\right).
\end{equation}
One can easily check that due to this phase convention 
\begin{eqnarray}
-i\sigma^{2}(h_{\hat{k},r})^{*}  =  -re^{-ir\phi}h_{\hat{k},-r}, \qquad 
h_{-\hat{k},r}  =  -h_{\hat{k},-r}.
\end{eqnarray}
Using the above relations, one obtains
\begin{equation}
V_{\vec{k},r}(\eta,\vec{x})=\frac{e^{-i\vec{k}\cdot\vec{x}}}{(2\pi)^{3/2}}\left(\begin{array}{c}
-u_{B,k,\eta}^{*}\\
ru_{A,k,\eta}^{*}
\end{array}\right)\otimes h_{-\hat{k},r}\cdot(e^{-ir\phi}).
\end{equation}
The normalization of the mode functions implies 
\begin{eqnarray}
h_{\hat{k},r}^{\dagger}h_{\hat{k},s}  =  \delta_{rs}, \qquad |u_{A,k,\eta}|^{2}+|u_{B,k,\eta}|^{2}  =  1.
\end{eqnarray}
With this ansatz, Eq.~(\ref{eq:rescaled_EOM}) simplifies as follows:
\begin{equation}
i\partial_{\eta}\left(\begin{array}{c}
u_{A}\\
u_{B}
\end{array}\right)=\left(\begin{array}{cc}
am & k\\
k & -am
\end{array}\right)\left(\begin{array}{c}
u_{A}\\
u_{B}
\end{array}\right).\label{eq:Dirac Equation}
\end{equation}
Let $\tilde{U}_{\vec{k},s}$ be another basis in the form of
Eq.~(\ref{eq:mode-ansatz}).  Due to the orthogonality of $h_{\hat{k},r}$ and
$e^{i\vec{k}\cdot x}$ , $\tilde{U}_{\vec{k},s}$ can only be a linear
combination of $U_{\vec{k},s},V_{-\vec{k},s}$:
\begin{equation}
\tilde{U}_{\vec{k},s}=\alpha_{(\vec{k},s)(\vec{k},s)}U_{\vec{k},s}+\beta_{(\vec{k},s)(-\vec{k},s)}V_{-\vec{k},s}
\end{equation}
The Bogoliubov coefficients are extracted using the scalar
product of the mode functions evaluated at time $\eta$ as follows:
\begin{eqnarray}
\alpha_{(\vec{k},s)(\vec{k},s)} & = & u_{A,k,\eta}^{*}\tilde{u}_{A,k,\eta}+u_{B,k,\eta}^{*}\tilde{u}_{B,k,\eta}\\
\beta_{(\vec{k},s)(-\vec{k},s)} & = &
e^{-is\phi(\hat{k})}(u_{A,k,\eta}\tilde{u}_{B,k,\eta}-u_{B,k,\eta}\tilde{u}_{A,k,\eta})
\label{eq:betak}
\end{eqnarray}
Since we will only consider $|\beta_{k}|^{2}$ in this work, we can drop the $e^{-is\phi(\hat{k})}$ factor in the $\beta_{k}$
definition without loss of generality.  Here one of the bases (corresponding to the
Heisenberg state of the universe) is specified by asymptotic conditions such as the Bunch-Davies boundary condition as the in-vacuum (see e.g.~Eq.~(\ref{eq:in-BC}).)  Similarly, the other basis is the observable operator basis as specified by asymptotic conditions at late times, which is referred to as the out-vacuum.

\section{Demonstration that $|\beta_{k}|^{2}\sim\frac{1}{2}$ for Small $k$ }
We begin with the determination of $\beta_k$ from Eq.~(\ref{eq:betak}) evaluated at very late
times when the out-modes can be directly replaced by their asymptotic values.  In the limit in which $am/k\rightarrow\infty$, we see that we then only need to find the asymptotic values of the in-modes:
\begin{eqnarray}
|\beta_{k}|  =  |u_{A,k,\eta}^{out}u_{B,k,\eta}^{in}-u_{B,k,\eta}^{out}u_{A,k,\eta}^{in}|= 
 |\sqrt{\frac{\omega+am}{2\omega}}u_{B,k,\eta}^{in}-\sqrt{\frac{\omega-am}{2\omega}}u_{A,k,\eta}^{in}|= 
  \lim_{\eta\rightarrow\infty}|u_{B,k,\eta}^{in}|.
\end{eqnarray}
 Let us consider the evolution equations as given in Eq.~(\ref{eq:Dirac Equation}) with boundary conditions as given in Eq.~(\ref{eq:in-BC}).  For concreteness, we choose a time $\eta_{i}$ that is early enough such that $u_{A}(\eta_{i})\approx u_{B}(\eta_{i})\approx\frac{1}{\sqrt{2}}$. The system can be formally solved  to obtain  
\begin{eqnarray}
\left(\begin{array}{c}
u_{A}\\
u_{B}
\end{array}\right)_{f} & = & T\exp\left\{ -i\int d\Phi\sigma(\theta)\right\} \left(\begin{array}{c}
u_{A}\\
u_{B}
\end{array}\right)_{i}
\end{eqnarray}
in which $\omega\cos\theta=k$, $\omega\sin\theta=am$,  $\omega d\eta=d\Phi$, and $\sigma(\theta)=\sigma_{1}\cos\theta+\sigma_{3}\sin\theta$ ($0 \leq \theta \leq \pi/2$). The time evolution is thus expressed as a series of infinitesimal
$SU(2)$ rotations that act successively on the complex vector $u\equiv ( u_A \,  u_B )$. 

 For fixed $\theta$, the evolution corresponds to precession about the axis defined by $\sigma(\theta)$.  However, throughout the evolution of the universe, $\sigma(\theta)$ evolves from its initial direction along $\sigma_1$ ($am\ll k$) to its final direction along $\sigma_3$ ($am \gg k$).   If the switching of the axis is much faster than the precession time scale, $u$ remains in the $xy$-plane and rotates around the new axis $\sigma_{3}$, while if the switching is much slower compared with the precession time scale, $u$  adheres closely to the rotation axis and thus ends up in the $\sigma_{3}$ direction.   The time scale of the axis switching is given by the Hubble expansion rate, since the universe needs to expand several e-folds for $am$ to overtake $k$, while the  time scale of the precession is given by the physical frequency $\omega/a$, which is on the order of $m$ during the transition.  Hence, fast transitions occur when $m\ll H$, for which $|u_{B}|^{2}$ stabilizes at $\frac{1}{2}$ and $|\beta_{k}|^{2}=\frac{1}{2}$.
After $H(\eta)$ drops below $m$, only slow transitions occur and $|\beta_{k}|^{2}$
is small. 

\section{\label{sec:heavymasscase} Heavy mass case ($m>H_{e})$}
As we expect the particle production spectrum $|\beta_{k}|^{2}$ to be
exponentially suppressed by $m/H$, we can adopt a similar approach as
the heavy mass scalar case \cite{Chung2003} to look for a one-pole
approximation to the time integral that determines $\beta_k$.  We shall
consider the time-dependent Bogoliubov coefficients between the
in-modes and the zeroth adiabatic modes with boundary conditions such
that
\begin{equation}
\left(\begin{array}{c}
u_{A}\\
u_{B}
\end{array}\right)_{k,\eta=\eta_{1}}^{(\eta_{1})}=\left(\begin{array}{c}
\sqrt{\frac{\omega+am}{2\omega}}\\
\sqrt{\frac{\omega-am}{2\omega}}
\end{array}\right).
\end{equation}
In the above, the superscript $(\eta_{1})$ indicates the time that the
boundary conditions are imposed. The in-modes can be decomposed into
the zeroth adiabatic mode basis as follows:
\begin{equation}
\left(\begin{array}{c}
u_{A}\\
u_{B}
\end{array}\right)_{k,\eta_{1}}^{in}=\alpha_{k}^{in-(\eta_{1})}\left(\begin{array}{c}
u_{A}\\
u_{B}
\end{array}\right)_{k,\eta_{1}}^{(\eta_{1})}+\beta_{k}^{in-(\eta_{1})}\left(\begin{array}{c}
-u_{B}^{*}\\
u_{A}^{*}
\end{array}\right)_{k,\eta_{1}}^{(\eta_{1})}.\label{eq:in-inst}
\end{equation}
For  $\eta_{1}\rightarrow\infty$, the instantaneous-modes will
coincide with the out-modes up to an overall phase, and hence
\begin{equation}
|\beta_{k}|=\lim_{\eta_{1}\rightarrow\infty}|\beta_{k}^{in-(\eta_{1})}|.
\end{equation}
Inserting this decomposition into Eq.~(\ref{eq:Dirac Equation}) (and
writing $\alpha_{k}^{in-(\eta_{1})}$ as $\alpha_{k}(\eta_{1})$,
etc.~for notational simplicity) results in
\begin{eqnarray}
\dot{\alpha}_{k}(\eta_{1}) =  -\frac{mk}{2\omega^{2}}\dot{a}e^{2i\int^{\eta_{1}}d\eta\omega(\eta)}\beta_{k}(\eta_{1}), \qquad
\dot{\beta}_{k}(\eta_{1}) =  \frac{mk}{2\omega^{2}}\dot{a}e^{-2i\int^{\eta_{1}}d\eta\omega(\eta)}\alpha_{k}(\eta_{1}),\label{eq:betak-wkb}
\end{eqnarray}
with the initial conditions
$\alpha_{k}(\eta_{i})=1,\,\beta_{k}(\eta_{i})=0$ for the time
$\eta_{i}$ early enough that the mode is inside the dS event
horizon.  Since we expect $|\beta_{k}|\ll1$ and $a_{k}\approx1$, we can
replace $\alpha=1$ in Eq.~(\ref{eq:betak-wkb}) and formally write the
solution as
\begin{equation}
\beta_{k}(\eta_{f})=\int_{\eta_{i}}^{\eta_{f}}d\tau\frac{mk}{2\omega^{2}}\dot{a}(\tau)e^{-2i\int^{\tau}d\eta\omega(\eta)}.
\label{eq:steepestdescentstart}
\end{equation}
The steepest descent method can be applied to evaluate this integral
in a similar fashion as was done for the scalar case in
\cite{Chung2003}.  Despite the different $k$ dependence in
Eq.~(\ref{eq:steepestdescentstart}), the result is the
same as Eq.~(41) of \cite{Chung2003}:
\begin{equation}
|\beta_{k}|^{2}\approx\exp\left\{ -4\left[\frac{[k/a(r)]^{2}}{m\sqrt{H^2(r)+R(r)/6}}+\frac{m}{\sqrt{H^2(r)+R(r)/6}}\right]\right\}, \label{eq:hm_result}
\end{equation}
in which $r$ is the real part of the complexified conformal time
$\tilde{\eta}$ at which $\omega(\tilde{\eta})=0$ and $R$ is the Ricci
scalar.  This is approximately
  due to the fact that the branch point occurs when $\omega=0$, such
  that the dominant contribution occurs when $\vert k/a \vert\sim m$.
Eq.~(\ref{eq:hm_result}) leads to the particle number density (fermion plus
anti-fermion) as
\begin{equation}
\rho_\psi(t) \approx \frac{1}{2
  \pi^{3/2}}\left(\frac{a(r)}{a(t)}\right)^3m\left[
  \frac{m}{4}\sqrt{H^2(r)+ R(r)/6} \right]^{3/2} \exp\left(\frac{-4
  m}{\sqrt{H^2(r)+R(r)/6}}\right).
\label{eq:heavymasscase}
\end{equation}
To estimate the relic abundance from this equation, one can use the
formula
\begin{equation}
\Omega_\psi h^2 \sim 100 \left(\frac{T_{\mbox{rh}}}{10^9
  \mbox{GeV}} \right)  \left(\frac{H(t_e)}{10^{13} \mbox{GeV}}
\right)^{-2}
\frac{\rho_\psi(t_e)}{\left(10^{12}\mbox{GeV}\right)^4},
\label{eq:largemassrelic}
\end{equation}
where one is only formally evaluating the $\rho_\psi(t_e)$ at the end
of inflation time $t_e$ even though the particle densities are well
defined at times far later than time.  Unlike the formulae presented
in the body of the text, the exponential sensitivity and the
approximations made in obtaining the saddle-point does not allow one
to guarantee an order of magnitude numerical accuracy, especially for
large $m/H(r)$ \cite{Chung2003}.  However, the spectral and mass
cutoffs can be well estimated by Eqs.~(\ref{eq:hm_result}) and
(\ref{eq:heavymasscase}).
\end{appendix}

\begin{acknowledgments}
This work is supported in part by the DOE through the grant
DE-FG02-95ER40896.  
\end{acknowledgments}

\bibliographystyle{JHEP}
\bibliography{particleprodbib}
\end{document}